\documentclass[10pt]{article}
\usepackage{graphicx}
\usepackage[square,sort&compress]{natbib}

\newcommand{\microG}{$\mu$G}
\def \deeg {$^\circ$}
\def \ebv {E(B-V)}
\def \kms {${\rm km~s}^{-1}$}
\def\cmtwo{cm$^{-2}$}
\def\cc{cm$^{-3}$}
\def\HI{H$^\mathrm{o}$}
\def\nHI{n(H$^\mathrm{o})$}
\def\nHII{n(H$^\mathrm{+})$}
\def\NHI{N(H$^\mathrm{o}$)}

\def\glon{$\ell$}
\def\glat{$b$}
\begin{document}
\begin{center} 
{\bf \Large Connecting the interstellar magnetic field at the heliosphere to the Loop I superbubble} 
\end{center}

\begin{center}
{P C Frisch$^1$, A Berdyugin$^2$, H O Funsten$^3$, A M Magalhaes$^4$, D J McComas$^5$, \\
V Piirola$^2$, N A Schwadron$^6$, D B Seriacopi$^4$, and S J  Wiktorowicz$^7$} \\
\begin{tiny}
\vspace*{0.02in}
{$^1$Dept of Astronomy and Astrophysics, University of Chicago, Chicago, IL 60637, USA} 
{$^2$Finnish Centre for Astronomy with ESO, University of Turku, Finland} 
{$^3$Los Alamos National Laboratory, Los Alamos, NM} 
{$^4$Inst de Astronomia, Geofisica e Ciencias Atmosfericas, Universidade de Sao Paulo, Brazil} 
{$^5$Southwest Research Inst, San Antonio, TX and University of Texas, San Antonio, TX} 
{$^6$Space Science Center, University of New Hampshire, Durham, NH}  
{$^7$Dept of Astronomy, University of California at Santa Cruz, Santa Cruz, CA}  
\end{tiny}
\end{center}

\setcounter{page}{1}

\begin{abstract}
The local interstellar magnetic field affects both the heliosphere and
the surrounding cluster of interstellar clouds (CLIC).  Measurements
of linearly polarized starlight provide the only test of the magnetic
field threading the CLIC.  Polarization measurements of the CLIC
magnetic field show multiple local magnetic structures, one of which
is aligned with the magnetic field traced by the center of the
``ribbon'' of energetic neutral atoms discovered by the Interstellar
Boundary Explorer (IBEX).  Comparisons between the bulk motion of the
CLIC through the local standard of rest, the magnetic field direction,
the geometric center of Loop I, and the polarized dust bridge
extending from the heliosphere toward the North Polar Spur direction
all suggest that the CLIC is part of the rim region of the Loop I
superbubble.
\end{abstract}

\section{Origin of local interstellar magnetic field}

The Sun resides in a region of space with very low interstellar
densities [\citenum{Frisch:2011araa}] that is offset by $\sim 400$ pc from
the density maximum of the Orion spiral arm (also known as the Orion
Spur, \citenum{Elias:2009}).  The interstellar magnetic field (ISMF) in
this interarm region is characterized by an ordered magnetic field
with a strength $1.4 \pm 0.3$ \microG\ that is directed toward the
local radiant at $\ell \sim 88^\circ$ in the galactic plane
[\citenum{RandKulkarni:1989,JinLinHan:2009,Heiles:1996Brev}].  The older
clusters of the Scorpius-Centaurus Association (ScoOB2), located $\sim
120$ pc from the Sun, are also offset from the density maximum of the
Orion arm [\citenum{Elias:2009}].

\begin{figure}[t!]
\hspace{1.1cm}
  \centering
  \includegraphics[width=4.5in]{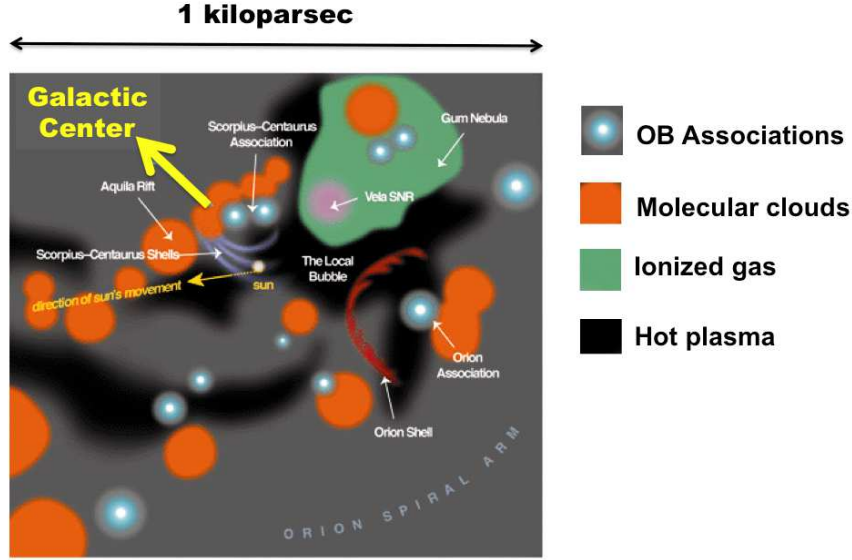}
\caption{\label{fig:apod}The distribution of molecular clouds (orange)
  traced by the CO 12 cm line is shown for nearby regions.
  Stellar associations (bright starry dots) form from
  these molecular clouds, generating superbubble shells that compress
  interstellar dust and gas during expansion (gray and red shell-like
  arcs).  The solar motion through the LSR (golden arrow) has carried
  the Sun through the low density interior of the Local Bubble for the
  past several million years (figure from [\citenum{Frisch:2000amsci}]).  }
\end{figure}

A superbubble was created by the ScoOB2 stellar winds and supernova
during three epochs of star formation over the past 15 Myrs
[\citenum{deGeus:1992,Crawford:1991sca,Frisch:1995rev,MaizApellaniz:2001}].
A classic expanding superbubble will sweep up the ISMF during
expansion, expanding asymmetrically in directions parallel to the ISMF
for initially uniform gas, and with the compressed magnetic field
lines parallel to the rim of the shell in directions perpendicular to
the magnetic field
[\citenum{Vallee:1984,MacLowMcCray:1988,FerriereZweibel:1991}]. The epochs
of star formation in ScoOB2 swept the expanding superbubble into the
interarm region through which the heliosphere moves
(Fig. \ref{fig:apod}, [\citenum{Frisch:1981,Frisch:1995rev,MaizApellaniz:2001}]).
The interstellar signatures of the Loop I superbubble have been
analyzed based on tracers of neutral gas, polarized starlight, and
radio continuum emission
[\citenum{Berkhuijsen:1971,Heilesetal:1980nps,deGeus:1992,Crawford:1991sca,Heiles:1998whence,Heiles:1998lb,Santos:2010,HartmannBurton:2012,BerdyuginPiirola:2014}].

The original superbubble expanded into a giant dust cloud, the
remnants of which form the Aquila Rift molecular cloud, and initiated
the sequential star formation that produced the upper Scorpius
subgroup [\citenum{deGeus:1992,Crawford:1991sca}]. Recent stellar
evolution in the upper Scorpius subgroup generated the supernova that
produced the bright X-ray source associated with the North Polar Spur
radio emission.  Observations and models of the X-ray plasma and radio
shell show that it is a supernova remnant with an age of $\sim 2$ Myrs
that was reheated possibly within the past 100,000 years
[\citenum{Iwan:1980npsXrayloopI}].  Hydrogen column densities and dust
opacity jump at $\sim 100$ pc in the North Polar Spur segment of Loop
I ($330^\circ < \ell < 30^\circ$, $0^\circ < b < 40^\circ$,
[\citenum{Iwan:1980npsXrayloopI,ReisCorradi:2008}]).  Cosmic rays
generated in the remnant light up the magnetic field compressed in the
superbubble shell with synchrotron emission, producing the giant
non-thermal radio continuum feature, Loop I, that dominates the
northern hemisphere sky [\citenum{Berkhuijsen:1971}].  The magnetic field
associated with Loop I dominates starlight polarization in the
northern hemisphere
[\citenum{MathewsonFord:1970,Berkhuijsen:1971,Santos:2010}], and
the polarized emission of interstellar dust grains
[\citenum{Planck:2014polz}].

\begin{figure}[t!]
\centering
\vspace*{-1.1in}
\includegraphics[width=4.5in]{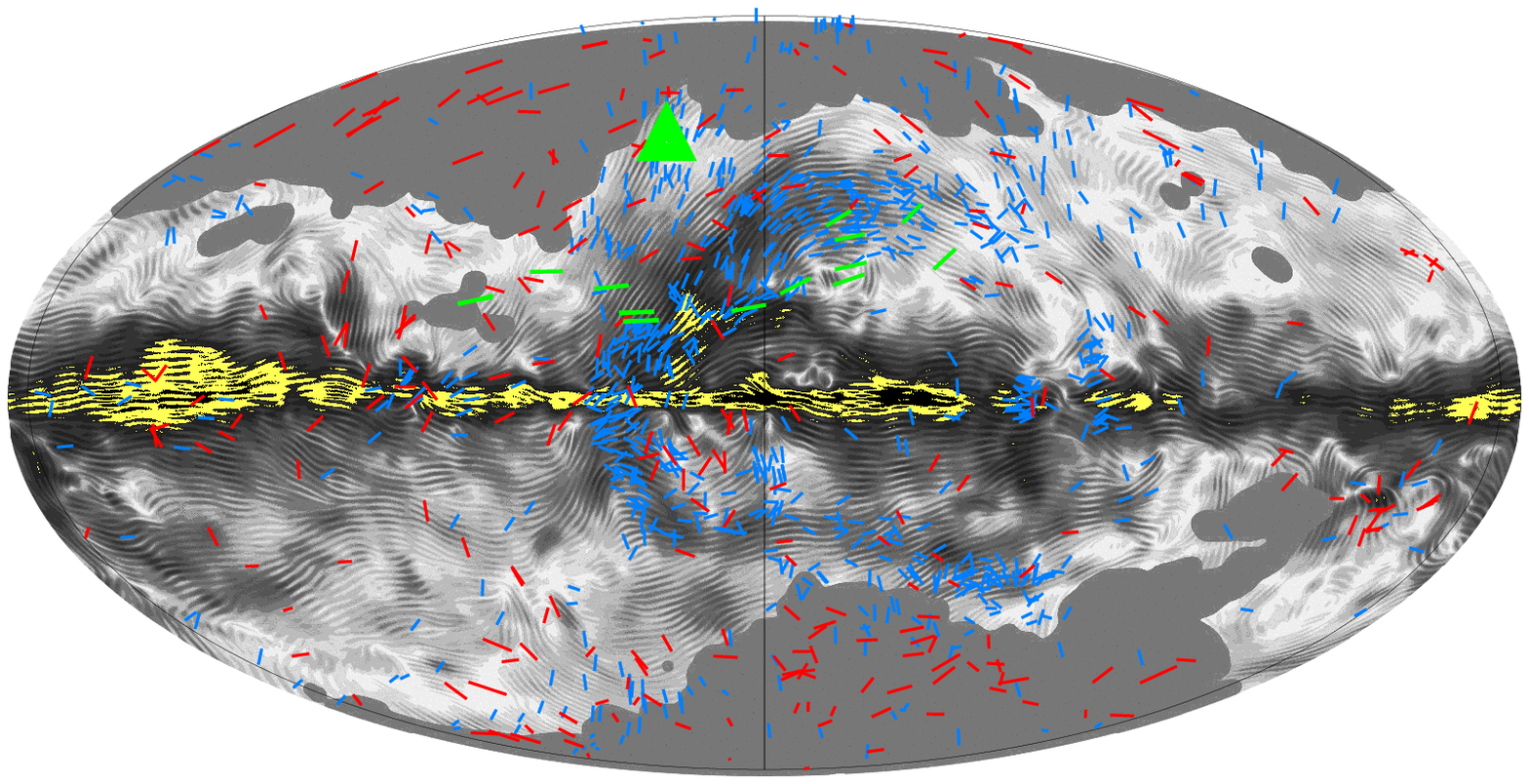}
\caption{\label{fig:planck} Stellar polarization data are overplotted
  on the magnetic field direction based on polarized dust emission
  measured by Planck \citep{Planck:2014polz}.  The green triangle
  shows the direction of the ISMF traced by the center of the IBEX
  Ribbon arc \citep{Funsten:2013}.  A high-latitude dust arc is found
  in this direction.  The blue bars show the polarizations of distant
  ($>40$ pc) stars that trace the maximum of the Loop I intensity.
  Red bars show polarizations for stars within 40 pc.  Green bars show
  the filament stars that are located toward the inner ridge of the
  nearby portion of Loop I (Fig. \ref{fig:ebv}).  In the northern
  parts of Loop I and in the first galactic quadrant,, where the ISMF
  extends to the solar location (Fig. \ref{fig:DvrP}), the ISMF
  direction traced by the optical and infrared polarizations are
  consistent.  Yellow regions appear in sightlines of highest dust
  emissivity.}
\end{figure}

A coherent picture of the Loop I ISMF can found using three sources of
polarized light, starlight polarization, polarized dust emission, and
polarized synchrotron emission.  Starlight is linearly polarized by
asymmetric dust grains that have larger opacities along the axis that
is perpendicular to the magnetic field, so that optical polarization
vectors are parallel to the magnetic field direction
(e.g. [\citenum{Lazarian:2007rev}).  The polarized emission of dust
grains is parallel to the most optically opaque axis, and therefore
perpendicular to the ISMF direction for aligned grains
[\citenum{Planck:2014polz,Planck:2014optical}].  The polarization of
synchrotron emission is aligned with the acceleration vector of
electrons and ions propagating along the ISMF, so that it is
perpendicular to the ISMF direction for galactic cosmic rays gyrating
around the ISMF.  The property that the optical polarization vectors
are perpendicular to the Loop I synchrotron polarizations
(e.g. compare polarization maps in references
[\citenum{MathewsonFord:1970}and [\citenum{Berkhuijsen:1971}]), and also
perpendicular to polarized infrared light from dust associated with
Loop I [\citenum{Planck:2014polz,Planck:2014optical}], shows that the
position angles of polarized starlight trace the magnetic field
directions in the global diffuse interstellar medium, and that the
linearly polarized starlight has a plane of polarization that is
parallel to the magnetic field direction.  This property enables the
mapping of the local ISMF direction using high-sensitivity
polarization measurements, $\leq 0.01$\%
[\citenum{planetpol:2010,Frisch:2010s1,Frisch:2010ismf1,Frisch:2012ismf2,Frisch:2014ismf3}].
Fig. \ref{fig:planck} shows starlight polarizations plotted over the
ISMF direction obtained from measurements of polarized dust emission
by Planck [\citenum{Planck:2014polz,Planck:2014optical}].  

\vspace*{0.1in}

Optically polarized starlight reveals the ISMF associated with Loop I
[\citenum{MathewsonFord:1970}].  Polarization surveys by Mathewson and
Ford [\citenum{MathewsonFord:1970}], Santos et al. [\citenum{Santos:2010}] and
Berdyugin et al. [\citenum{BerdyuginPiirola:2014}] have mapped starlight
polarizations that trace the ISMF of Loop I.  Dust extinction and
polarization strengths jump at $100 \pm 20$ pc where $\ell <
40^\circ$, and at $280 \pm 50$ pc where $360^\circ < \ell <
270^\circ$, indicating that the densest parts of Loop I extend closest
to the Sun for longitudes $<40^\circ$.

\section{Location of the heliosphere in the rim of Loop I} 

\begin{figure}[t!]
\centering
\includegraphics[width=4.5in]{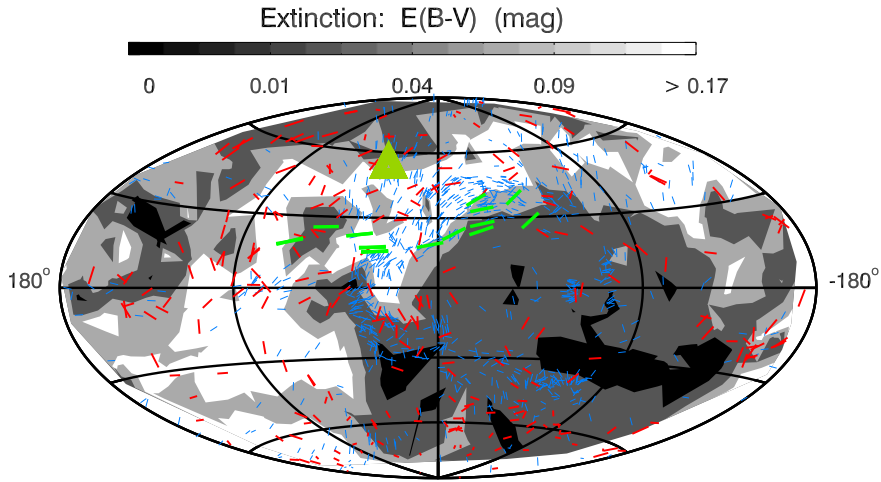}
\vspace*{-0.5in}
\caption{\label{fig:ebv} Smoothed contours of color excess \ebv\ for
  stars 50--100 pc show that dust associated with the inner rim of
  Loop I extends to inside of 100 pc in the North Polar Spur region.
  Optical polarization vectors for stars within 40 pc (red) show
  similar position angles in some regions when compared to more
  distant stellar polarizations (blue) that trace the Loop I ISMF.
  The filament stars are plotted in green.  The green triangle
  identifies the IBEX Ribbon ISMF direction.  The void in ISM in the
  fourth galactic quadrant forms where the interiors of the Loop I
  superbubble and Local Bubble merge.  The plot is centered on
  $\ell=0$, and galactic longitude increases to the left.  Figure
  details are given in \citep{Frisch:2012ismf2}.  }
\end{figure}

The Loop I regions of highest column density are located beyond $\sim
100$ pc, but spatially smoothed \ebv\ data show that the Loop I dust
shell is apparent within 50--100 pc at high latitudes and in the first
galactic quadrant (Fig. \ref{fig:ebv}).  The void in the dust that is
centered near $\ell \sim 320^\circ$, $b \sim -20^\circ$ shows that the
Loop I superbubble interior merges into the Local Bubble interior in
the fourth galactic quadrant ($\ell=270^\circ - 360^\circ$).  The only
interstellar cloud found near the center of the void is the G-cloud
[\citenum{RLIV:2008}], where the color excess corresponding to the low
observed column densities (\ebv$<0.001$ mag) is too small for
detection.

Several types of data suggest that the Sun is embedded in the rim of
the evolved Loop I superbubble.  The first is that the geometry of the
arcs associated with Loop I places the Sun at the bubble edge.
Assuming that synchrotron arcs and \HI\ arcs associated with the North
Polar Spur ($\ell \sim 10^\circ - 30^\circ$) form a single superbubble
places the center of Loop I at $\ell=320^\circ$, $b=12^\circ$,
distance 120 pc, with a radius of $\sim 115$ pc
[\citenum{BerkhuijsenHaslam:1971,Heiles:1998lb}].  An alternative
geometric model is consistent with the series of high-latitude arcs
visible in the \HI\ data
[\citenum{Heilesetal:1980nps,Heiles:1998whence}]. The associated radio
continuum arcs have been modeled as two shells by Wolleben
[\citenum{Wolleben:2007}], using the intensity of polarized radio
continuum emission (also see [\citenum{BerdyuginPiirola:2014}]) and
Fig. 6).  The shell geometry suggests that the Sun is most likely in
the rim of the shell identified as ``S1'' [\citenum{Wolleben:2007}], which
is centered at $\ell = 346^\circ \pm 5^\circ$, $b = 3^\circ \pm
5^\circ$
[\citenum{Wolleben:2007,Frisch:2010s1,Frisch:2012ismf2,FrischSchwadron:2013icns}].

The kinematical properties of nearby interstellar gas provide the
second type of information suggesting that the Sun is located in the
rim of Loop I.  Interstellar gas within 30 pc flows past the Sun with
an upwind direction in the Local Standard of Rest (LSR) that is
directed toward the center of the Loop I superbubble
[\citenum{FGW:2002,Frisch:2011araa}].  For the assumption that the
different cloudlets in the flow move with a rigid body motion, the
heliocentric vector flow of nearby ISM past the Sun is away from $\ell
= 12.4^\circ$, $b = 11.6^\circ $, at a velocity of $-28.1 \pm 4.6$
\kms\ [\citenum{FGW:2002,Frisch:2011araa}].  Using the LSR motion derived
from Hipparcos data by Schonrich et al. (2010), and including the
uncertainties of both the solar apex motion and the heliocentric
interstellar motion, this heliocentric vector flow corresponds to the
upwind LSR velocity vector of \glon,\glat$=338.4^\circ \pm 15.2^\circ,
-5.3^\circ \pm 8.3^\circ$ and $17.2 \pm 4.6$ \kms\ for the
bulk CLIC motion.  This is $11^\circ \pm 19^\circ$ from the S1 shell
center.  Alternate velocity vectors for local gas yield similar
results.  For example, a local cloud velocity derived from a
restricted data set that omits outlying and redundant velocity
components produced a heliocentric upwind velocity of $\ell =
5.8^\circ \pm 0.8^\circ$, $b = 12.8^\circ \pm 0.7^\circ $, velocity
$-25.5 \pm 0.3$ \kms\ [\citenum{GryJenkins:2014clic}], which then
corresponds to the LSR velocity $\ell = 323.6^\circ \pm 1.7^\circ$, $b
= -16.2^\circ \pm 1.4^\circ $ and a velocity of $-17.0 \pm 0.6$ \kms.
Thus, regardless of the specific assumptions about the individual
components used for calculating the bulk flow, the upwind direction is
always toward the low density void around the Loop I center.
\nocite{BinneyDehnen:2010}

\begin{figure}[t!]
\centering
\includegraphics[width=4.5in]{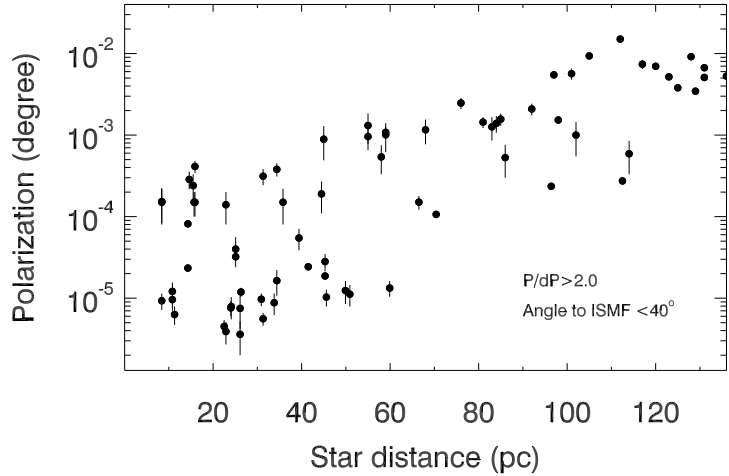}
\caption{\label{fig:DvrP}Polarizations are plotted against the
  distances of stars that are within $40^\circ$ of the magnetic field
  direction indicated by the center of the IBEX Ribbon (see text).
  The steady increase in polarization with distance shows that the
  aligned dust grains extend to the solar location.  The closest
  interstellar polarizations are toward stars within 5 pc, indicating
  that the stream of polarized dust toward Loop I must extend close to
  the heliosphere in the first galactic quadrant.}
\end{figure}

The third set of data relating the CLIC with the perimeter of Loop I
is plotted in Fig. \ref{fig:DvrP}, where the polarization strengths
are shown to increase steadily with distance inside of a cone of
radius $40^\circ$ that is centered on the IBEX ribbon ISMF direction
(also see related figures in [\citenum{planetpol:2010}] and
[\citenum{Frisch:2012ismf2}).  The IBEX ribbon ISMF is directed toward
the North Polar Spur (Figs. \ref{fig:planck}, \ref{fig:ebv}), so this
polarization bridge shows that both an ISMF and a dust stream extend
from the solar location out to the interstellar density jump at 100 pc
in the North Polar Spur direction.  The connection that we find
between the ISMF direction at the heliosphere and the dominant local
field direction from the polarization data
[\citenum{Frisch:2012ismf2,Frisch:2014ismf3}] then links the very local
ISMF to the ISMF of Loop I.

This cluster of local interstellar cloudlets (CLIC) is decelerating,
as shown by blue-shifted velocity components in both the upwind and
downwind directions [\citenum{FGW:2002,Frisch:2011araa}].  Representative
of this deceleration is the supersonic collision between the LIC and
the BC in the direction of Sirius
[\citenum{Schwadron:2014sci,FrischSchwadron:2013icns}].  

\section{Interstellar magnetic field shaping the heliosphere}

The discovery of a ``ribbon'' of energetic neutral atoms (ENAs) by the
Interstellar Boundary Explorer (IBEX) spacecraft
[\citenum{McComas:2009sci,Funsten:2009sci,Schwadron:2009sci}] has provided
a remarkable opportunity to understand the relation between the
dominant magnetic structure in the sky, Loop I, and the ISMF that
shapes the heliosphere.  Optical polarization data that trace the ISMF
in the low-density local interstellar medium (ISM) gives us a new tool
for probing the physical properties and history of nearby interstellar
clouds.  We have undertaken a project to map the local ISMF direction,
and are finding that the ISMF that shapes the heliosphere is
associated with the rim of the Loop I superbubble
[\citenum{Frisch:2010ismf1,Frisch:2012ismf2,Frisch:2014ismf3}].

Although there is no generally accepted physical mechanism for the
formation of the IBEX ribbon [\citenum{McComasLewisSchwadron:2014}], most
existing models indicate that it appears in sightlines that are
perpendicular to the ISMF as it drapes over the heliosphere upwind of
the heliopause [\citenum{Schwadron:2009sci}].  The ribbon is
extraordinarily circular, with an ellipticity of $<0.3$, and is
centered at galactic coordinates of $\ell = 34.7^\circ$,
$b=56.6^\circ $ ($\pm 2.6^\circ$, after converting the J2000 ecliptic
position in [\citenum{Funsten:2013}] to galactic coordinates).  The class
of models that have successfully reproduced the ribbon configuration
predict that the ribbon forms $\sim 50 $ AU upwind of the heliopause,
and that the ISMF direction is within $\sim 15^\circ $ of the center
of the ribbon arc [\citenum{Heerikhuisen:2014}].  Because of the
importance of interstellar pressure terms to the shape and dimensions
of the outer heliosphere, the geometric configuration of the ribbon is
highly sensitive to the physical properties of the surrounding
interstellar cloud
[\citenum{Heerikhuisen:2014,HeerikhuisenPogorelov:2011,Frisch:2010next,Ratkiewicz:2012ribbon}].

The polarity of the local ISMF is not defined close to the Sun by
astronomical data.  It has been measured by Voyager 1 in the
heliosphere depletion region, and the field was found to be directed
upwards out of the ecliptic plane [\citenum{Burlaga:2013ismf}].
Observations of the Faraday rotation measures of four pulsars in the
third galactic quadrant region of low interstellar densities also find
that the polarity of the ISMF over $\sim 100 $ pc scales is directed
upwards through the galactic and ecliptic planes [\citenum{Salvati:2010}].

\section{Structure in the local interstellar magnetic field}

We have determined the magnetic field direction in the local
interstellar medium where gas column densities are low,
\NHI$<10^{18.7}$ \cmtwo, dust is unreddened, \ebv$<0.001$ mag, and
interstellar optical polarizations are weak, $<0.01$\%
[\citenum{Frisch:2010ismf1,Frisch:2012ismf2,Frisch:2014ismf3}].  These
studies use new polarization data collected in the southern and
northern hemispheres as well as data in the literature.  The present
analysis focuses on data for stars within 40 pc and 90\deeg\ of the
heliosphere nose in order to confine the study to local ISM in the
upwind hemisphere of the sky.

We have developed a method [\citenum{Frisch:2010ismf1,Frisch:2012ismf2,Frisch:2014ismf3}]
for obtaining the ISMF direction that takes advantage of the fact that
a linear polarization position angle of zero degrees indicates that
the starlight polarization vector is aligned with the ISMF.  Testing
the data sample for all possible field directions then yields the
``true'' field direction corresponding to the lowest value of the
weighted mean of the sine of those position angles.  The weighting
factors allow the use of lower sensitivity polarization data.  The
polarization data used for this study are plotted in
Fig. \ref{fig:ebv} as the red and green polarization vectors.  The
polarizations plotted in green consist of a distinct group of stars,
here termed the ``filament stars'' (see below).  
\begin{figure}
\begin{center}
\includegraphics[width=4in]{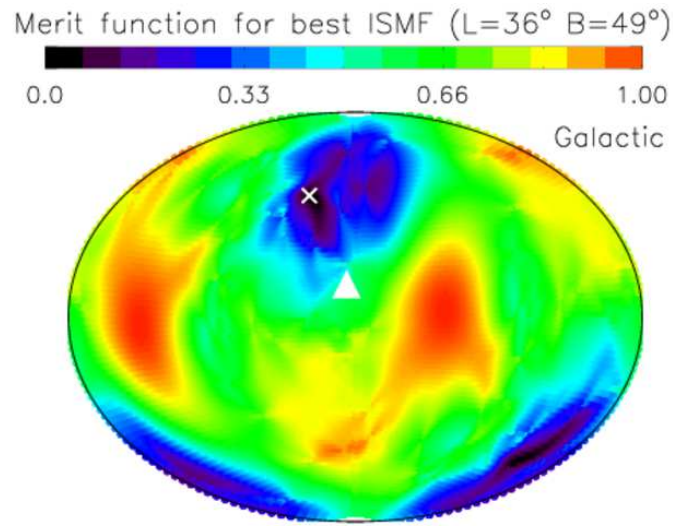}
\end{center}
\caption{\label{fig:meritf} The spatial variations of the merit function for the best-fitting ISMF direction obtained
from our polarization data set of qualifying stars (e.g.within 40 pc and 90\deeg\ of the heliosphere nose)
are plotted in galactic coordinates, using an Aitoff projection centered on the galactic center.
The white "X" and triangle indicate the ISMF and heliosphere nose directions, respectively.
The polarization data set underlying this figure omits the filament stars that trace a different
magnetic field direction (see \citep{Frisch:2014ismf3} for details ).
The best-fitting ISMF direction is obtained by minimizing a merit function, which
in essence minimizes the ensemble of polarization position angle sines. 
The best-fitting ISMF direction to the polarization data is within $7.6^\circ \pm 16.2^\circ$ of the magnetic
field direction indicated by the IBEX ENA Ribbon \citep{Funsten:2013}.
}
\end{figure}

Frisch et al. [\citenum{Frisch:2012ismf2}] applied this analysis method to
a polarization dataset and found a best-fitting magnetic field
direction toward $\ell = 47^\circ \pm 20^\circ$, $b = 25^\circ \pm
20^\circ$.  The angle between this ISMF direction and the IBEX ribbon
field direction is $32^\circ \pm 28^\circ$.

In our new study [\citenum{Frisch:2014ismf3}], using an expanded dataset,
we find that at least two distinct magnetic field directions are
traced by the polarization data [\citenum{Frisch:2014ismf3}].  The first
structure is traced by the group of stars termed here the `filament
stars'; it is well-defined both by the elongated spatial grouping
(green polarizations in Fig. \ref{fig:ebv}), and by the linear
decrease of the polarization position angles with distance.  This
magnetic structure is sampled by stars as close as $\sim 5$ pc.  More
details and possible origins of this filament are given in
[\citenum{Frisch:2014ismf3}].

When the filament stars are omitted from the calculations of the
best-fitting ISMF, the ISMF that dominates the fit has the direction
$\ell = 36^\circ $, $b = 49^\circ$ ($\pm 16^\circ$).  The values of
the merit function for this fit are plotted for magnetic field
directions in all positions on the sky (Fig. \ref{fig:meritf}).  This
ISMF direction represents the dominant magnetic structure in the
hemisphere around the heliosphere nose, according to our fitting
algorithm.  The direction is in excellent agreement with the ISMF
direction from the IBEX ribbon center to within the uncertainties.
The angle between this dominant ISMF direction and the LSR velocity
vector of the bulk flow of the local ISM past the Sun is $74^\circ \pm
24^\circ$.  A nearly perpendicular angle between the CLIC flow
velocity and the CLIC magnetic field is consistent with the expected
configuration of a magnetic field that has been compressed in an
expanding superbubble shell.

The strength of the ISMF can not be obtained from optical polarization
data, but other data and models indicate that it is $\sim 3$ \microG.
If an ISMF strength of 3 \microG\ is assumed in the Heerikhuisen et
al.  models of the ribbon [\citenum{Heerikhuisen:2014}], the ribbon center
is offset from the direction of the ISMF at infinity by $\sim
4^\circ$. The offset becomes larger for weaker field strengths.
Photoionization models of the Local Interstellar Cloud (LIC) around
the heliosphere find a field strength of $\sim 3$ for equilibrium
between magnetic and thermal pressures
\microG\ [\citenum{SlavinFrisch:2008}].  The pressure of the plasma in the
inner heliosheath traced by IBEX ENA measurements, compared to
interstellar pressures, indicates a strength for the ISMF of $\sim
3.3$ \microG\ [\citenum{Schwadronetal:2011sep}].

The ISMF is a conduit for TeV galactic cosmic rays that flow into the
heliosphere [\citenum{Schwadron:2014sci}]. TeV cosmic rays diffuse through
the local ISM with typical gyroradii of less than 700 AU.  Stellar
polarization data show that the ISMF traced by the IBEX ribbon extends
into the space beyond the immediate heliosphere vicinity.  This
supports results showing that the observed TeV cosmic ray anisotropies
are consistent with cosmic ray diffusion along the ISMF traced by the
IBEX ribbon [\citenum{Schwadron:2014sci}].  
\begin{figure}[h!]
\begin{center}
\includegraphics[width=4.5in]{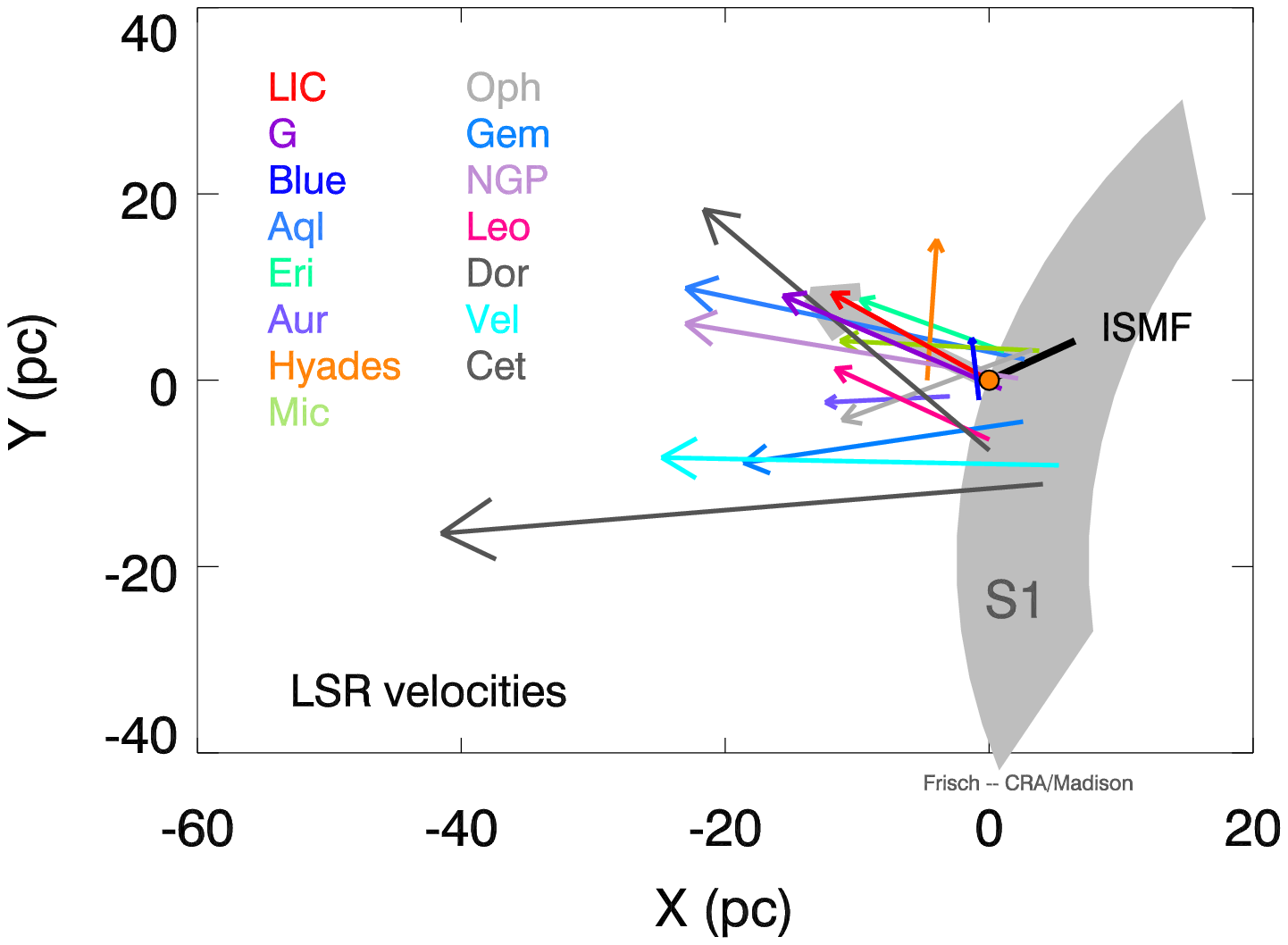}
\end{center}
\caption{\label{fig:s1} The Loop I S1 shell \citep{Wolleben:2007} and
  nearby cloud \citep{RLIV:2008} LSR velocities
  \citep{FrischSchwadron:2013icns} are shown projected onto the
  galactic plane.  The black line shows the ISMF direction from the
  IBEX ENA Ribbon center \citep{Funsten:2013}.  Omitting filament
  stars from the calculation gives an ISMF direction from the
  polarization data and IBEX Ribbon that agree.  The large gray arrow
  shows the bulk flow of local ISM through the LSR
  \citep{Frisch:2011araa}.  This ISM flow is perpendicular to the the
  ISMF and S1 shell.  The Sun is located at the orange dot.}
\end{figure}

The geometrical relation between the local magnetic field, the
positions and kinematics of local interstellar clouds, and the Loop I
S1 superbubble, suggest that the Sun is located in the boundary of
this evolved superbubble.  The quasi-perpendicular angle between the
bulk kinematics and magnetic field of the local ISM indicates that a
complete picture of low density interstellar clouds needs to include
information on the interstellar magnetic field.  Fig. \ref{fig:s1}
shows the projection of the Wolleben S1 shell [\citenum{Wolleben:2007}]
onto the galactic plane, with the bulk motion of the ISM flowing past the
Sun overplotted as the thick light gray arrow, and the fifteen clouds
derived in the Redfield and Linsky [\citenum{RLIV:2008}] analysis plotted
as individual color-coded arrows. The S1 shell geometry is consistent
with the inclusion of the CLIC within the shell.

\section{What is an interstellar cloud?}

Eugene Parker once asked ``what is an interstellar cloud''
[\citenum{Frisch:2001b}].  The low density partially ionized interstellar
gas in which the heliosphere is now embedded (\nHI$\sim 0.2$ \cc,
\nHII$\sim 0.06$ \cc, T$\sim 6300$ K [\citenum{SlavinFrisch:2008}]) was
once identified as ``intercloud medium'' [\citenum{RogersonYork:1973}] to
distinguish it from the dusty or opaque clouds that attenuate
starlight.  Beyond about 7 AU, the mass density of interplanetary
space is dominated by neutral interstellar hydrogen. In some sense
this interstellar material forms a heliospheric interstellar cloud.

Interstellar material can be characterized by volume density, column
density, ionization, kinematics, turbulence, temperature, composition,
and/or the magnetic field.  The tradition of identifying interstellar
clouds by velocity began with an early all-sky survey of interstellar
absorption components toward bright stars where multiple
Doppler-shifted components were resolved [\citenum{Adams:1949}].  Since
then, the Doppler-shifted velocity of a cloud has proven the easiest
form of cloud identification.  The shortcoming of flagging clouds by
absorption line velocity is that this diagnostic depends on the
instrument.  Adjacent absorption components typically are blended in
velocity space, and the number of identified clouds has been found to
increase nearly exponentially as instrumental spectral resolution
improves [\citenum{WeltyK:2001,Frisch:2001b}].  The question of ``what is
a cloud'' then becomes even more problematical for the decelerating
flow of ISM past the Sun.

This tradition of identifying clouds by velocity has been followed for
the local ISM, but the concept that clouds are individual entities is
transcended by the inclusion of an interstellar magnetic field.  We
have found that the bulk flow of local interstellar gas past the Sun
has a direction that is nearly perpendicular to the direction of the
local ISMF determined from optical polarization data
[\citenum{Frisch:2012ismf2,Frisch:2014ismf3,FrischSchwadron:2013icns,Schwadron:2014sci}]).
This connection between cloud kinematics and magnetic field extends
the concept of interstellar clouds beyond velocity components, so as
to include large scale structural features in the ISMF such as the
Loop I superbubble.  Furthermore, the ISMF shaping the heliosphere
corresponds to the ISMF traced by the polarizations of nearby
starlight.

A discussion of the multiple models for the ``cloud-structure'' of the
local ISM is outside of the scope of this paper.  However the
comparison between the polarization data (e.g. Fig. \ref{fig:ebv}]) and
the 15-cloud structure modeled by Redfield and Linsky
([\citenum{RLIV:2008}], also see [\citenum{Frisch:2011araa}]) produces the
interesting result that no significant nearby polarizations are yet
found in the upwind direction toward the G-cloud.  This appears to be
related to the destruction of grains by interstellar shocks
[\citenum{Frisch:2014ismf3}].  The relative velocities between adjacent
cloud components support the presence of local shocked ISM, both
between the LIC and G-cloud
[\citenum{FrischYork:1991,GrzedzielskiLallement:1996}], the Blue Cloud and
the LIC [\citenum{Schwadron:2014sci}], and other nearby sightlines.

\section{Conclusions}
Measurements of nearby interstellar gas and dust indicate that the
local ISM is associated with the rim of the evolved Loop I
superbubble.  The bulk motion of nearby interstellar gas through the
LSR indicates an upwind direction toward the center of Loop I.  Both
polarized starlight and the center of the IBEX ribbon indicate that
the magnetic field is perpendicular to the bulk velocity of local ISM.
The ISMF compressed in the rim of an expanding superbubble shell that
is sweeping up interstellar material is expected to be perpendicular
to the expansion velocity of the shell.  The increase of polarizations
with distance indicates that the source of some of the CLIC dust is
part of an interstellar dust stream that connects to the nearest side
of Loop I toward the North Polar Spur region.

{\bf Acknowledgements:} To be published in the proceedings of the 13th
Annual International Astrophysics Conference: ``Voyager, IBEX, and the
Interstellar Medium''.  This work has been partially supported by the
NASA Explorer Program through funding of the IBEX project.


\newcommand\araa{ARA\&A}
\newcommand\mnras{MNRAS}%
\newcommand\ssr{Space~Sci.~Rev.}%
\newcommand\nat{Nature}%
\newcommand\memras{MmRAS}%
\newcommand\jqsrt{J.~Quant.~Spec.~Radiat.~Transf.}%
\newcommand\apj{ApJ}
\newcommand\apjl{ApJ}
\newcommand\apjs{ApJS}
\newcommand\aap{A\&A}

\end{document}